\documentclass{jfm}

\usepackage{graphicx}
\usepackage{epstopdf, epsfig}
\usepackage{graphicx}
\usepackage{dcolumn}
\usepackage{bm}
\usepackage{color}
\usepackage{array}
\usepackage{amsmath,amssymb,stmaryrd} 
\usepackage{hyperref}
\usepackage{soul} 

\definecolor{red}{rgb}{0.8500, 0.1250, 0.0480} 

\ifCUPmtlplainloaded \else
  \checkfont{eurm10}
  \iffontfound
    \IfFileExists{upmath.sty}
      {\typeout{^^JFound AMS Euler Roman fonts on the system,
                   using the 'upmath' package.^^J}%
       \usepackage{upmath}}
      {\typeout{^^JFound AMS Euler Roman fonts on the system, but you
                   dont seem to have the}%
       \typeout{'upmath' package installed. JFM.cls can take advantage
                 of these fonts,^^Jif you use 'upmath' package.^^J}%
      }
  \else
  \fi
\fi

\ifCUPmtlplainloaded \else
  \checkfont{msam10}
  \iffontfound
    \IfFileExists{amssymb.sty}
      {\typeout{^^JFound AMS Symbol fonts on the system, using the
                'amssymb' package.^^J}%
       \usepackage{amssymb}%

      }{}
  \fi
\fi

\ifCUPmtlplainloaded \else
  \IfFileExists{amsbsy.sty}
    {\typeout{^^JFound the 'amsbsy' package on the system, using it.^^J}%
     \usepackage{amsbsy}}
    {\providecommand\boldsymbol[1]{\mbox{\boldmath $##1$}}}
\fi

\title[Network Structure of Two-Dimensional Decaying Isotropic Turbulence]
{Network Structure of Two-Dimensional Decaying Isotropic Turbulence} 

\author[K. Taira, A. G. Nair, and S. L. Brunton]%
{
Kunihiko Taira$^1$\thanks{Email address for correspondence: ktaira@fsu.edu},
Aditya G. Nair$^1$, and Steven L. Brunton$^2$
}

\affiliation{$^1$Department of Mechanical Engineering, 
Florida State University, Tallahassee, FL 32310, USA\\ [\affilskip]
$^2$Department of Mechanical Engineering, 
University of Washington, Seattle, WA 98195, USA}

\pubyear{2016}
\volume{???}
\pagerange{???--???}
\begin{document}

\maketitle

 
\begin{abstract}
The present paper reports on our effort to characterize vortical interactions in complex fluid flows through the use of network analysis.  In particular, we examine the vortex interactions in two-dimensional decaying isotropic turbulence and find that the vortical interaction network can be characterized by a weighted scale-free network.  It is found that the turbulent flow network retains its scale-free behavior until the characteristic value of circulation reaches a critical value.  
Furthermore, we show that the two-dimensional turbulence network is resilient against random perturbations but can be greatly influenced when forcing is focused towards the vortical structures that are categorized as network hubs.  These findings can serve as a network-analytic foundation to examine complex geophysical and thin-film flows and take advantage of the rapidly growing field of network theory, which complements ongoing turbulence research based on vortex dynamics, hydrodynamic stability, and statistics.  While additional work is essential to extend the mathematical tools from network analysis to extract deeper physical insights of turbulence, an understanding of turbulence based on the interaction-based network-theoretic framework presents a promising alternative in turbulence modeling and control efforts.
\end{abstract}



\section{Introduction}
\label{sec:intro}

Recent advances in the field of network analysis have revealed the structures of internet, technological, social and biological networks \citep{Albert:RMP02, Newman:SIAMReview03, Barrat:PNAS04, Newman10}.  Having characterized these networks, we are able to study the dynamics such as disease outbreak and information propagation on networks and analyze resilience of network-based activities \citep{Barrat08, Albert:Nature00}.  These analysis techniques are founded on graph theory, dynamical systems, and operator theory but place unique emphasis on interactions and connectivity amongst the elements that establish a network.  Thus far, most of the applications of network analysis have been concerned with discrete settings in which nodes are individualized quantities, such as people, organisms, equipment, or stations \citep{Caldarelli07, Newman10}.  In this paper, we extend the network analysis to continuous representation of physical phenomena, in particular two-dimensional turbulence.

The chaotic motion of a large number of vortices in turbulent flows is caused by the induced velocities of the vortices themselves.  What makes turbulence rich and complex are the vortical interactions in the flow field that take place over a wide range of length scales \citep{Tennekes72, Hinze75, Frisch95, Pope00, davidson2004turbulence, Lesieur08}. Thus, complete understanding of turbulence has remained a challenge to this day because of its high-dimensionality, multi-scale interactions, nonlinearity and the resulting chaos.  Network science provides an alternative view of complex fluid flows in terms of a network of vortex interactions \citep{Nair:JFM15}, and this perspective illuminates the underlying structure and organization of turbulent flows.  In this work, we show that two-dimensional isotropic turbulence \citep{Kraichnan:RPP80, McWilliams:JFM84, Benzi:PRA90, Benzi:PF92, davidson2004turbulence, Boffetta:ARFM12} has a scale-free network structure reminiscent of other networks found in nature \citep{Barabasi:Science99, Caldarelli07}.  While most of the attention has been placed on unweighted scale-free networks, we consider the use of weighted scale-free network to describe the variations in the strength of interactions or connectivities \citep{Barrat:PRL04}.   Upon identifying the network structure of turbulence, physical insights can be obtained as to which vortical interactions are important in capturing the overall physics and how it may be possible to control the dynamics of turbulent vortices \citep{Liu:Nature11, Farazmand:JFM11, Brunton:AMR15}.

\section{Problem description and approach}

\begin{figure}
\begin{center}
\includegraphics[width=0.4\textwidth]{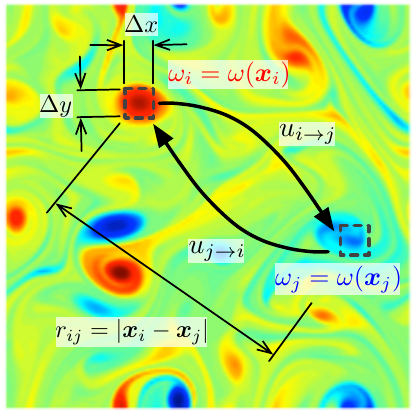}
\end{center}
\caption{Interaction of fluid elements in two-dimensional turbulence.  The strength of the vortical interaction between elements $i$ and $j$ having vorticity $\omega_i$ and $\omega_j$ is quantified through the induced velocities $u_{i\rightarrow j}$ and  $u_{j \rightarrow i}$, respectively.  For discretizing the Cartesian domain, we take $n_x$ and $n_y$ points in the horizontal and vertical directions, respectively, providing the adjacency matrix $\boldsymbol{A}$ of size $n\times n$ with $n = n_x n_y$.  Shown in the background with a contour plot is the corresponding vorticity field with initial $Re(t_0) = 814$ at $t = 18$.}
\label{fig:setup} 
\end{figure}

To extract the network structure of the flow, we quantify the interactions between fluid elements based on the vortical interactions.  The velocity $\boldsymbol{u}$ at position $\boldsymbol{x}$ induced by the vorticity distribution $\boldsymbol{\omega}$ of the flow is 
\begin{equation}
   \boldsymbol{u}(\boldsymbol{x},t) 
   = \frac{1}{4\pi} 
   \int \frac{\boldsymbol{\omega}(\tilde{\boldsymbol{x}},t) \times (\boldsymbol{x}-\tilde{\boldsymbol{x}})}
   {|\boldsymbol{x} - \tilde{\boldsymbol{x}}|^3} d\tilde{\boldsymbol{x}}.
   \label{eq:vel3d}
\end{equation}
In this study, we focus on unforced two-dimensional isotropic turbulence in a periodic box and assess the influence of the vorticity distribution over a Cartesian domain.  Here, the two-dimensional vorticity field reduces to $\boldsymbol{\omega}(\boldsymbol{x},t) = \omega(x,y,t)\hat{\bf e}_z$ with $\hat{\bf e}_z$ denoting the unit normal plane vector.  Modeling the vortical component for each discrete Cartesian element as a line vortex, we can evaluate how fluid elements influence each other, as depicted in Fig.~\ref{fig:setup}.  
Here, the magnitude of the induced velocity from fluid element $i$ on another element $j$ reduces from Eq.~(\ref{eq:vel3d}) to
\begin{equation}
   u_{i \rightarrow j} = \frac{|\gamma_i|}{2\pi | \boldsymbol{x}_i-\boldsymbol{x}_j |}, 
   \label{eq:vel2d}
\end{equation}
where $\gamma_i = \omega(\boldsymbol{x_i}) \Delta x \Delta y$ is the circulation of fluid element $i$ with side lengths of $\Delta x$ and $\Delta y$.  The superposition of the induced velocity from all other fluid elements provides the advective velocity of the fluid element.  Detailed discussions on using point vortices to develop the network-theoretic framework for describing unsteady vortical flows can be found in \cite{Nair:JFM15}.  Note that adjacency matrices are commonly defined with positive weights as considered here, but they can be relaxed to accommodate positive and negative weights within the context of vortical interactions.  This point will be revisited later.

To assess and describe the vortical interactions in the flow field, we utilize a weighted network (graph).  The definition of a network (graph) $\mathcal{G}$ requires sets of vertices (nodes) $\mathcal{V}$, edges $\mathcal{E}$, and weights $\mathcal{W}$ \citep{Newman10}.  With these three components defined, a graph can be uniquely determined, i.e., $\mathcal{G} = \mathcal{G}(\mathcal{V}, \mathcal{E}, \mathcal{W})$.  The nodes $\mathcal{V}$ in this study are taken to be the vortical elements residing within the Cartesian cells and the edges $\mathcal{E}$ represent the vortical interactions between those vortical elements.  The edge weights $\mathcal{W}$ quantify the strengths of the vortical interactions.   Given $n$ nodes, a collection of the weights $w_{ij}$ in the form of a matrix $\boldsymbol{A} \in \mathbb{R}^{n \times n}$ with
\begin{equation}
   A_{ij} = \begin{cases} w_{ij} & \text{if~} (i,j) \in \mathcal{E} \\ 0 & \text{otherwise}, \end{cases}
\end{equation}
is called the adjacency matrix and is used to describe the network connectivity.  In the above definition, $A_{ij}$ is set to the edge weight $w_{ij}$ if there exists an edge (interaction) between nodes $i$ and $j$.  Details on the fundamental concepts involved in network theory can be found in \cite{Newman10} and \cite{Dorogovtsev10} with descriptions of vortical-interaction networks in \cite{Nair:JFM15}.

Based on Eq.~(\ref{eq:vel2d}), we define the network adjacency matrix as the average induced velocity
\begin{equation}
   A_{ij} = 
   \begin{cases} \frac{1}{2} ( u_{i \rightarrow j} + u_{j \rightarrow i}) & 
   \text{if~} i\ne j \\ 0 & \text{otherwise}
   \end{cases}
   \label{eq:adj}
\end{equation}
to quantify the magnitude of interaction between fluid elements $i$ and $j$ \citep{Nair:JFM15}.  Note that an element cannot impose velocity upon itself, which is captured by the null entry along the diagonal of the adjacency matrix.  In the present study, the influence from the neighboring periodic vortex images are also accounted for in the analysis.  This formulation yields a full matrix except for its diagonal entries that are identically zero.  In assessing the strength of the vortical interaction between two fluid elements, we utilize Eq.~(\ref{eq:adj}) to perform network analysis to extract the spatial connectivity structure.  This approach has been successful in capturing the nonlinear vortex dynamics and modeling the trajectories of vortex clusters \citep{Nair:JFM15}.  The adjacency matrix considered here is symmetric to quantify the average interaction strength.  Note that the geometric mean can be alternatively chosen and yields similar results.  In general, the adjacency matrix can be formulated in an asymmetric manner:
\begin{equation}
   A_{ij} = \begin{cases} \phi  u_{i \rightarrow j} + (1-\phi) u_{j \rightarrow i} & \text{if~} i\ne j \\ 0 & \text{otherwise}. \end{cases}
   \label{eq:adj_asymmetric}
\end{equation}
Here the parameter $\phi$ takes a value between $0$ and $1$.  For the aforementioned symmetric formulation in Eq.~(\ref{eq:adj}), $\phi$ is selected as $1/2$.  When $\phi = 0$ and $1$, the adjacency matrix $A_{ij}$ are defined by the velocity imposed to the other elements ($A_{ij} = u_{j \rightarrow i}$) and upon themselves ($A_{ij} = u_{i \rightarrow j}$), respectively, for $i\ne j$.  We mainly focus on the use of the symmetric adjacency matrix in this work but will consider the asymmetric formulation briefly to highlight the difference from a physical point of view in the next section.  We note in passing that the theoretical tools for symmetric adjacency matrices are more widely available compared to the asymmetric matrices.

The flow field analyzed in this study is obtained from direct numerical simulation on a square bi-periodic computational domain $(x,y) \in [0,L] \times [0,L]$ with a grid size of $m_x \times m_y = 1024 \times 1024$.  The unforced two-dimensional incompressible isotropic turbulent flow is simulated by numerically solving the two-dimensional vorticity transport equation 
\begin{equation}
   \frac{\partial \omega}{\partial t} + u_j \frac{\partial \omega}{\partial x_j} = \frac{1}{Re} \frac{\partial^2 \omega}{\partial x_j \partial x_j},
\end{equation}
where $\boldsymbol{u}$ and $\omega$ are the velocity and vorticity variables, respectively.  The simulation is performed with the Fourier spectral method and the fourth-order Runge--Kutta time integration scheme \citep{Canuto88}.  The vorticity field is initialized with a smooth distribution comprised of a large number ($\approx 100$) of superposed vortices \citep{Taylor:1918} with random strengths, core sizes, and locations chosen such that the kinetic energy spectra satisfies $E(k) \propto k \exp(-k^2/k_0^2)$, where $k_0 = 26.5$, following the setup by \cite{Brachet:PRL86} and \cite{Kida:JPSJ85}.  The initial core sizes are selected to be sufficiently small compared to the size of the computational domain \citep{McWilliams:JFM84} arranged in random positions.  The velocity variable is normalized by the square root of the spatial average of the initial kinetic energy $u^*(t_0) \equiv [\overline{u^2}(t_0)]^{1/2}$, where the overline denotes the spatial average.  The spatial length and time scales are non-dimensionalized by the initial integral length scale $l^*(t_0) \equiv [2\overline{u^2}(t_0)/\overline{\omega^2}(t_0)]^{1/2}$ and the initial eddy turnover time $t^*_0 \equiv l^*(t_0)/u^*(t_0)$, respectively.  The Reynolds number is defined accordingly as $Re \equiv u^* l^*/\nu$ where $\nu$ is the kinematic viscosity.  In this study, turbulent flows with initial Reynolds numbers of $Re(t_0) = 75$, $439$, $814$, $1607$, and $2485$ are selected.

\section{Results}

\subsection{Network-based characterization}

\begin{figure}
\begin{center}
\includegraphics[width=0.9\textwidth]{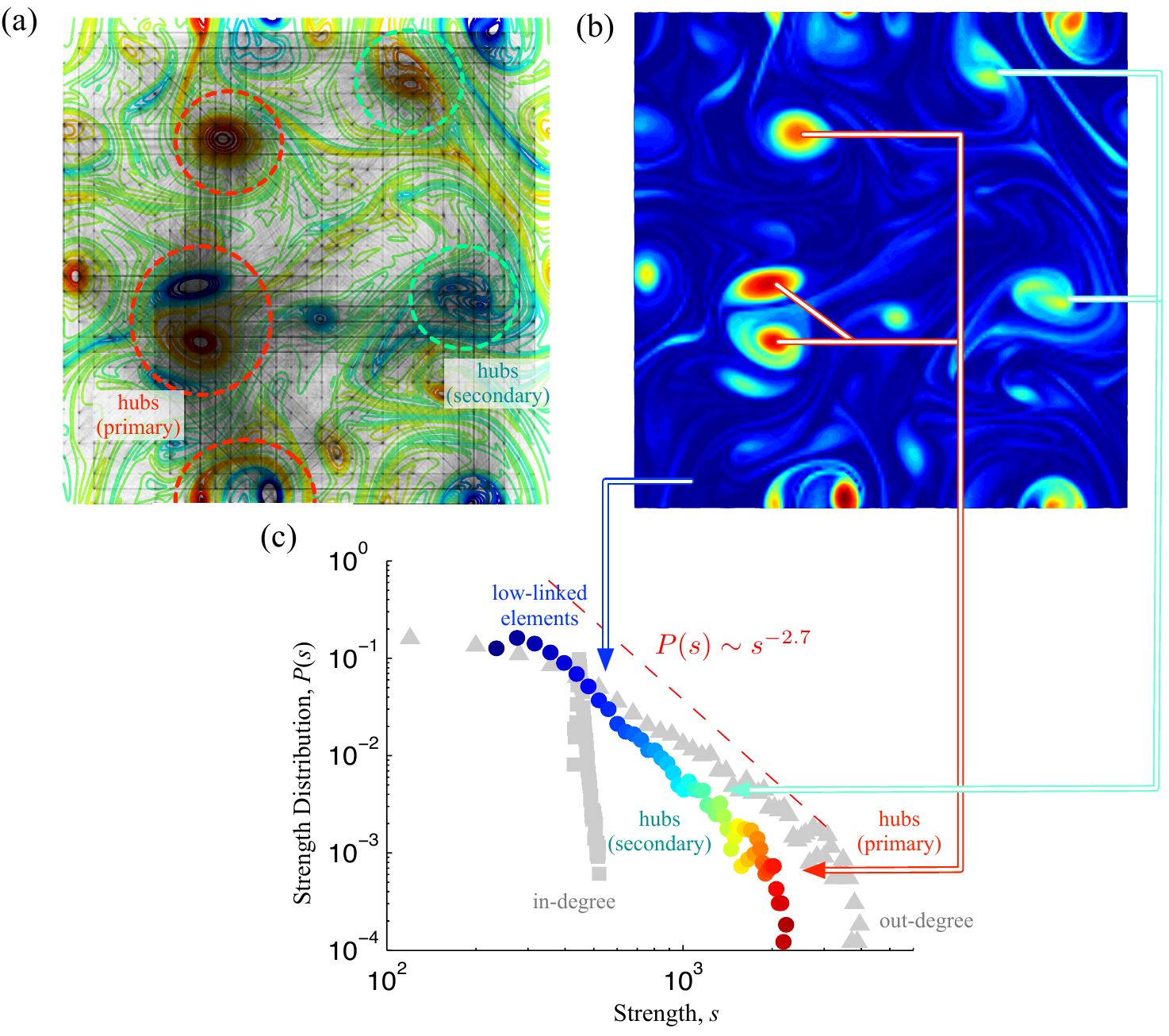}
\end{center}
\vspace{0.0in}
\caption{The scale-free network of vortical interactions in two-dimensional turbulence with initial $Re(t_0) = 814$.  (a) Turbulent network structure overlaid on the vorticity field with the darkness of the network edges corresponding to the values of the adjacency weights ($t = 18$).  (b) Contour plot of the node strength $s$ distribution.  Vortex cores having high degree of connectivity act as hubs in the turbulent vortical network. (c) The corresponding node strength probability distribution exhibiting the scale-free characteristics with $P \sim s^{-2.7}$.  The same contour level is shared by (b) and (c).  Also shown in the background of (c) in gray are the out and in-degree distributions ($\phi = 0$ and $1$, respectively).   The network visualized in (a) does not show interactions from periodic images and uses $32 \times 32$ nodes for graphical clarity.}
\label{fig:network} 
\end{figure}

We identify the underlying network structure and characteristics of two-dimensional turbulence based on the aforementioned symmetric adjacency weights.  The time-evolving vorticity field is obtained from a two-dimensional incompressible bi-periodic direct numerical simulation \citep{Canuto88} for unforced isotropic turbulence.  Given the vorticity field over a Cartesian grid, each fluid element is considered to be connected to all other elements through vortical network edges.  The resulting fluid flow network can in fact be described by a complete graph with a range of weights.  Next, we visualize the network edges with transparent gray scale corresponding to the adjacency weight, as shown in Fig.~\ref{fig:network}(a).  The captured structure reveals the turbulent network.  Some regions in the flow have a large number of strong connections corresponding to larger stronger vortices seen in red, serving as primary network hubs.  Note that these strong vortices induce velocities over long distances.  Moderate size vortices that act as secondary hubs also possess dominant connections to primary hubs and other secondary hubs.  In contrast, fluid elements corresponding to smaller, weaker eddies, shown in blue, generally have influence only in their vicinity.  The node strength distribution ($s_i = \sum_{j} A_{ij}$) over space shows that the vortices with large circulation have larger strength, as illustrated in Fig.~\ref{fig:network}(b).  The node strength distribution over space enables us to distinguish secondary and primary hubs, which may not be easily differentiated from simply visualizing the vorticity field or the $Q$ criterion in a traditional manner.  For instance, see the green vortices in (b) which can appear similar to primary ones in vorticity level.  

Plotting the probability of strength distribution $P(s)$ over the strength $s$ of fluid elements in Fig.~\ref{fig:network}(c), we find that two-dimensional isotropic turbulence network has a power-law distribution $P(s) \sim s^{-\gamma}$ with $\gamma = 2.7$ at the time shown.  This tells us that the vortex interactions in turbulence can be characterized by a weighted scale-free network.  This realization enables the interaction-based analysis of turbulent flows from a new perspective through network theory \citep{Newman10, Cohen10}.  In particular, this type of network is known to have certain resilience properties as we will explore later in this section.  Also shown in Fig.~\ref{fig:network}(c) in gray are the degree distributions for asymmetric adjacency formulations.  The out and in-degree distributions can be found by setting $\phi = 0$ and $1$, respectively, in Eq.~(\ref{eq:adj_asymmetric}).  It can be observed that the scale-free symmetric distribution is mostly comprised of the out-degree components, which describe how each vortical element influences all other elements (i.e., $u_{j \rightarrow i}$).  In contrast, we find that the in-degree distribution has a single peak which conveys that all fluid elements receive a similar amount of collective influence from vortices in the flow field.  We have found that the scale-free property of two-dimensional isotropic turbulence is most well-captured by the symmetric weights compared to the other asymmetric formulations.  It is also possible to examine the strength distribution taking positive and negative values of circulations, as we have briefly discussed in Section 2.  Utilizing positive and negative weights, their strength distribution can also exhibit a scale-free behavior but with network strength having both negative and positive values.  This leads to a symmetric strength distribution over the strength with resemblance to the probability density function of scaled displacements \citep{Weiss:PF98}.  In what follows, results based on the symmetric adjacency matrix (using the magnitude of induced velocity) are presented.

\begin{figure}
\begin{center}
\includegraphics[width=0.95\textwidth]{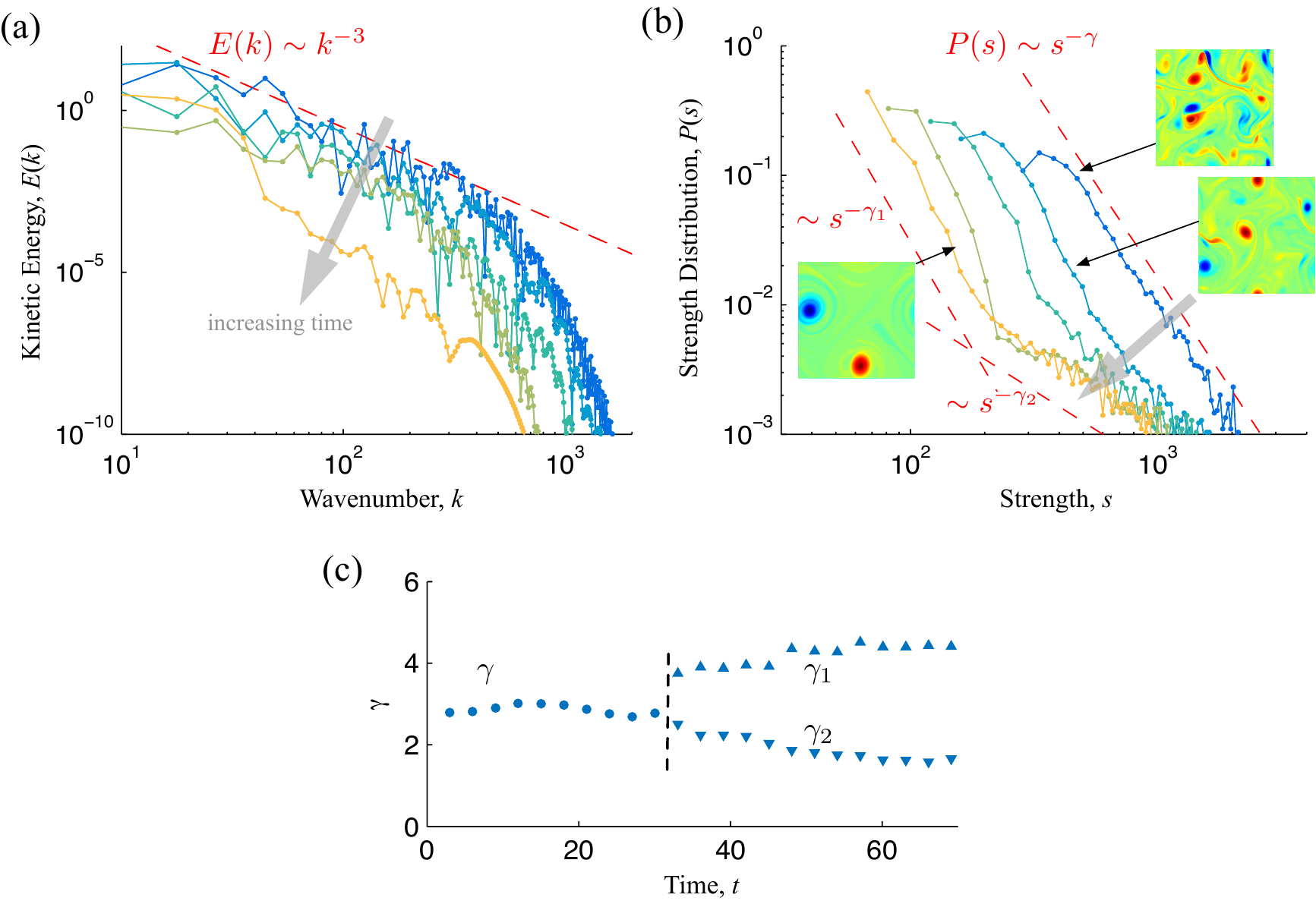} 
\caption{The dynamics of turbulent network with $Re(t_0) = 814$.  (a) Kinetic energy and (b) strength distribution of two-dimensional isotropic turbulence for $t = 15$, $30$, $75$, $150$, and $300$ (line colors represent time).  The inset plots in (b) show the corresponding vorticity fields.  The kinetic energy $E(k)$ is shown over the wavenumber $k$ exhibiting the asymptotic profile of $E(k) \sim k^{-3}$.   The strength distribution $P(s)$ displays the scale-free property of $P(s) \sim s^{-\gamma}$ over node strength $s$.  (c) The corresponding exponents $\gamma$, $\gamma_1$, and $\gamma_2$ are shown.  Later in time the strength distribution exhibits the emergence of two distributions, $P(s) \sim s^{-\gamma_1}$ and $s^{-\gamma_2}$.}
\label{fig:dynamics}
\end{center}
\end{figure}

Let us further examine the time-varying properties of the turbulent network.  In unforced turbulence, the kinetic energy of the flow decreases over time due to viscous dissipation as shown in Fig.~\ref{fig:dynamics}(a).  The strength distribution $P(s)$ of the turbulence network and the corresponding flow field snapshots are presented in Fig.~\ref{fig:dynamics}(b).  Turbulent flow is comprised of vortical structures over a wide range of spatial scales initially.   The distribution $P(s)$ exhibits scale-free characteristics with $P(s) \sim s^{-\gamma}$, where $\gamma \approx 2.7$, when the kinetic energy spectra exhibits the $k^{-3}$ profile for $t \lesssim 30$.  For the flow under consideration, a bend in the strength distribution appears for $t \gtrsim 30$ as the system starts to exhibit scale separation and loses the $k^{-3}$ energy spectra.   This is caused by the diffusion of smaller scale structures and their merging with other structures.  Over time, viscous dissipation removes kinetic energy through the smaller eddies and leaves only the larger vortices.  This behavior can be described by two power laws $P(s) \sim s^{-\gamma_1}$ and $P(s) \sim s^{-\gamma_2}$, where they capture the weaker fluid elements and the larger stronger vortices, respectively.  The bifurcation of these power laws is shown in Fig.~\ref{fig:dynamics}(c) indicated by the vertical dashed line.  We note that regardless of the initial condition used, the turbulent interaction network maintains the scale-free behavior in the present investigation as long as the energy spectra relaxes to the $k^{-3}$ profile \citep{Benzi:PRA90, Brachet:PRL86, Kida:JPSJ85}.  This scale-free behavior may be observed during the initial transient but is not a guaranteed common feature without the $k^{-3}$ energy spectra being present.

\begin{figure}
\begin{center}
\includegraphics[width=0.55\textwidth]{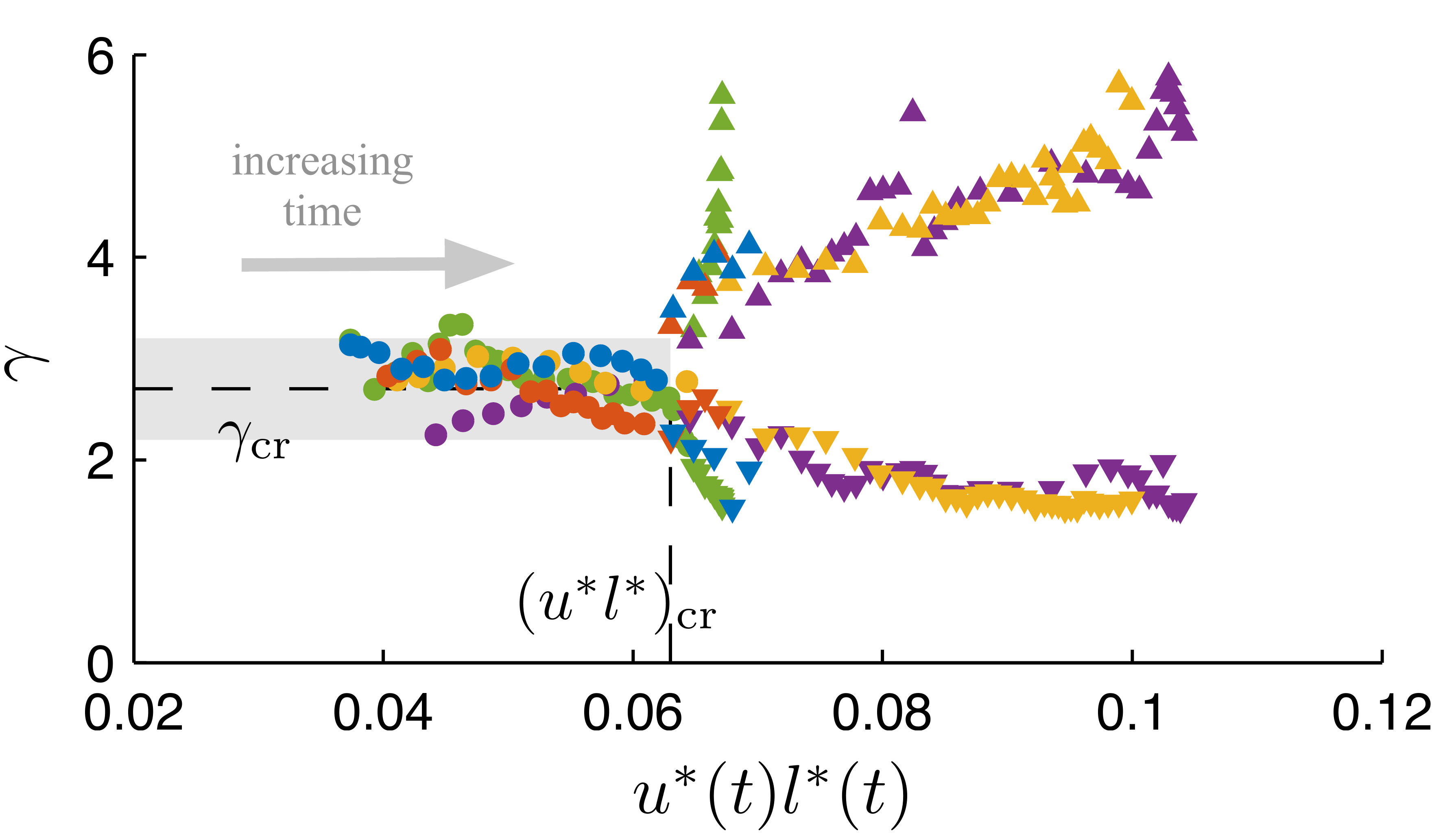} 
\caption{Exponent $\gamma$ for the network strength distribution $P(s) \sim s^{-\gamma}$ plotted over $u^*(t) l^*(t)$ with different initial Reynolds numbers (green: $Re(t_0) = 75$, purple: $Re(t_0) = 439$, yellow: $Re(t_0) = 814$, red: $Re(t_0) = 1607$, and blue: $Re(t_0) = 2485$).  Scale-free distributions are observed with $\gamma$ coalescing to $\gamma_\text{cr} \approx 2.7$ up until a bifurcation at $(u^* l^*)_\text{cr} \approx 0.063$.  The gray box shows $\gamma = -2.7 \pm 0.5$ as reference.}
\label{fig:gamma}
\end{center}
\end{figure}

We have considered a range of Reynolds numbers and observed that $\gamma$ takes values of $\gamma = 2.7 \pm 0.5$.  The variations observed in $\gamma$, $\gamma_1$, and $\gamma_2$ shown in Fig.~\ref{fig:dynamics}(c) are influenced by the chaotic nature of turbulence.  These parameters however appear to exhibit a coalescing behavior when they are plotted over the product of the characteristic velocity and length, $u^*(t) l^*(t)$.  Here, we interpret $u^*(t) l^*(t)$ as the circulation of vortices that have the characteristic velocity and length scales.  As shown in Figure \ref{fig:gamma}, we observe that the turbulence network shows coalescence of the scale-free parameter $\gamma$ to $\gamma_\text{cr} \approx 2.7$ over time for different cases of turbulent flows.  Once the flows reach a state where the characteristic strength of vortices is $( u^* l^* )_\text{cr} \approx 0.063$, the network distribution bifurcates to display two different slopes with $\gamma_1$ and $\gamma_2$, as previously illustrated in Figure \ref{fig:dynamics}.  This observation reveals that a scale-free turbulent network is present until the unforced turbulent flow field loses the smaller-scale vortices and mostly contains vortices with strengths larger than the critical value of $( u^* l^* )_\text{cr}$. 

\subsection{Resilience of turbulence networks}

Characterizing turbulent flow with a scale-free network enables us to view turbulent interactions in a systematic manner and provides insights into how vortical structures influence each other.   It is known from network analysis that scale-free networks are resilient to random perturbations but attacks towards network hubs can affect network dynamics in a detrimental manner \citep{Albert:Nature00}.  Network resilience for fluid flow translates to the difficulty of modifying the vortical interaction network and, consequently, the collective behavior of the vortices over time.   To measure the change in vortical interaction caused by network disturbance, we can consider how the removal of turbulence network nodes (percolation) modifies the characteristic network length 
\begin{equation}
   l_\text{network} \equiv \frac{1}{n(n-1)} \sum_{i\ne j} \min d(i,j),
\end{equation}
which is the average shortest network distance $d(i,j)$ between any two nodes on a network.  Here, we perform node percolation by setting the vorticity values at the chosen nodes to be zero.  The above metric quantifies how well vortical elements are connected within a turbulent network.  Note that the distance here refers to network distance based on the adjacency matrix and not the spatial distance.  In particular, we take the inverse of each adjacency weight $1/a_{ij}$ and evaluate the minimal sum 
\begin{equation}
   d(i,j) = 1/a_{ik_1} + 1/a_{k_1 k_2 } + \dots + 1/a_{k_m j}
\end{equation} 
over a network path that connects nodes $i$ and $j$ for this metric \citep{Rubinov:NI10}.  This metric $l_\text{network}$ can be thought of as the average of the minimal characteristic advective (commute) time per unit length between every pair of fluid elements in the domain.  This minimal network distance is determined using the Floyd--Warshall algorithm \citep{Floyd:ACM62}.

\begin{figure}
\begin{center}
\includegraphics[width=0.5\textwidth]{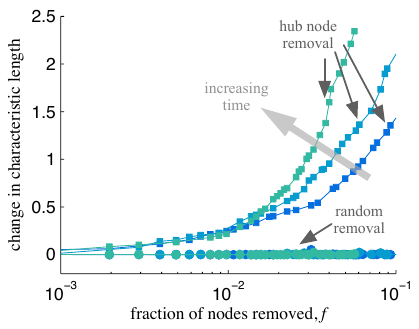} 
\caption{The resilience of turbulence network against node removals for $t = 15$, $30$, and $75$ with $Re(t_0) = 814$.  Shown are relative changes in the characteristic network length $\Delta \tilde{l}_\text{network}$ of turbulent flow for random node and hub node removals.  The colors of the curves represent the time when node removal is considered and follows Figure \ref{fig:dynamics}.}
\label{fig:resilience}
\end{center}
\end{figure}

The changes in the turbulence network characteristic length $l_\text{network}$ when network nodes are removed in a random fashion and a coordinated manner targeting hub nodes are summarized in Fig.~\ref{fig:resilience}.   Here, the changes in the normalized characteristic network length
\begin{equation}
   \Delta \tilde{l}_\text{network} \equiv \frac{l_\text{network}(t,f)-l_\text{network}(t,f=0)}{l_\text{network}(t,f=0)}
\end{equation}
for varied fraction of node removal are shown.  While it would be difficult to completely remove nodes as we have performed in this investigation, the present analysis sheds light on how external forcing or perturbations can alter the turbulent flow from an interaction-based analysis.
 We observe that turbulent flow is resilient against random forcing, as evident from the characteristic network length being unaffected even for a large fraction $f$ of nodes being removed.  This behavior is consistently observed over time.  On the other hand, we find that the global vortical interaction network can be greatly modified by targeting large vortex cores (hubs), as exhibited by the substantial change in the characteristic length.  It may be more energetically expensive to remove well-connected hub nodes, which often correspond to regions of concentrated vorticity. However, it is clear from Figure \ref{fig:resilience} that even the smallest fraction of hub node removal can influence the overall interaction, which suggests that hub removal still provides a more effective and efficient way to modify the flow than random node removal.
 
 When the vortical interaction network is grossly altered, the dynamics of the collection of vortices would be significantly modified \citep{Nair:JFM15}.  These observations also agree with past studies in flow control that identified effective actuation frequencies to be associated with the length scale of the large coherent structures in turbulent flows \citep{Joslin09, GadElHak00}.  With increasing time, we can further notice that network connectivity decreases with hub removal due to viscous dissipation of smaller vortical structures and the influence of removing the core structures becomes more evident.  The present network based understanding reveals which type of flow structures should be targeted with flow control if we aim to alter the behavior of the turbulent flow field in a global manner.   

\section{Concluding remarks}

The approach presented in this paper is the initial effort in performing network-based analysis of complex turbulent flows.  Using the mathematical toolsets from network-theory, we have identified that the vortical interactions in two-dimensional decaying isotropic turbulence have a scale-free network structure.  We have been able to reveal the structure by taking a continuous representation of the flow field and quantifying the network using a Cartesian discretization.  For two-dimensional isotropic turbulence, the node strength distribution was uncovered to be $P(s) \sim s^{-\gamma}$, where $\gamma = 2.7 \pm 0.5$.  Furthermore, we have found that the unforced turbulent flow field possesses an underlying scale-free network structure until the circulation of vortices with characteristic velocity and length scales reach $(u^* l^*)_\text{cr} \approx 0.063$.  By noticing that the turbulence network has scale-free characteristics, we were able to systematically show that the turbulence network is resilient against random perturbations but vulnerable against coordinated forcing on the hub vortices.  It should be noted that estimating and controlling each and every vortical structure in a turbulent flow is most likely improbable and impractical.  Instead, network analysis may provide a refreshing view point on how one can predict and modify the collective dynamics of vortices in the turbulent flow fields.  We believe that network-based analysis and control \citep{Mesbahi10, Liu:Nature11, Cornelius:NatureComm13, Kaiser:JFM14, Yan:NaturePhysics15} will provide novel mathematical fabric for paving the path towards network-based modeling and control of turbulent flows, which can potentially impact a wide spectrum of problems.


\section*{Acknowledgements}

K.T. and A.G.N. acknowledge the support from the US Army Research Office (Grant: W911NF-14-1-0386, Program manager: Dr.~Samuel Stanton) and the US Air Force Office of Scientific Research (Grant: FA9550-13-1-0183, Program manager Dr.~Douglas Smith).  S.L.B. acknowledges support by the Department of Mechanical Engineering at the University of Washington and as a Data Science Fellow in the eScience Institute.  


\bibliographystyle{jfm}
\bibliography{Taira_all}

\end{document}